\documentclass[review, 12pt]{elsarticle}

\usepackage{xcolor}
\usepackage{graphicx}
\usepackage{amssymb}
\usepackage{subfigure}
\usepackage{mathrsfs, amsmath}
\usepackage{footmisc}
\usepackage[margin=2.5cm]{geometry}

\journal{}

\newcommand{\markthis}[1]{#1}

\newcommand{\slideVel}{\ensuremath{V_0}}

\begin{document}

\begin{frontmatter}

\title{Distinct stick-slip modes in adhesive polymer interfaces}

\author{Koushik Viswanathan}
\address{Center for Materials Processing and Tribology, Purdue University, West Lafayette, IN}
\author{Narayan K. Sundaram}
\address{Indian Institute of Science, Bangalore, India}
\cortext[cor2]{Corresponding author: Koushik Viswanathan (kviswana@purdue.edu) }

\begin{abstract}
Stick-slip, manifest as intermittent tangential motion between two solids, is a well-known friction instability that occurs in a number of natural and engineering systems. In the context of adhesive polymer interfaces, this phenomenon has often been solely associated with Schallamach waves, which are termed slow waves due to their low propagation speeds. We study the dynamics of a model polymer interface using coupled force measurements and high speed \emph{in situ} imaging, to explore the occurrence of stick-slip linked to other slow wave phenomena. Two new waves---slip pulse and separation pulse---both distinct from Schallamach waves, are described. The slip pulse is a sharp stress front that propagates in the same direction as the Schallamach wave, while the separation pulse involves local interface detachment and travels in the opposite direction. Transitions between these stick-slip modes are easily effected by changing the sliding velocity or normal load. The properties of these three waves, and their relation to stick-slip is elucidated. We also demonstrate the important role of adhesion in effecting wave propagation. 
\end{abstract}


\end{frontmatter}


\section{Introduction}
When sliding two solids against one another, an interesting form of intermittent motion---stick-slip---is commonly observed, despite the application of a constant sliding force/ velocity  \cite{BowdenTabor_Friction_1973}. This motion is characterized by repeating cycles of stationary (stick) and moving (slip) phases at the interface. From the time it was first highlighted \cite{BowdenLeben_ProcRoySocA_1939}, stick--slip has emerged as the cause for a number of intermittent motions in nature. Examples include mechanical vibrations, squeals in automobile brakes \cite{Rabinowicz_SciAm_1956} and shallow earthquakes in the earth's crust \cite{BraceByerlee_Science_1966}. In addition, sustained stick--slip motion is also known to cause enhanced wear in polymers \cite{BartenevLavrentiev_FrictionWearPolymers} as well as in the degradation of human joints \cite{LeeETAL_ProcNatAcadSci_2013}. A detailed understanding of stick--slip dynamics is therefore of significant practical interest.

The following conceptual picture is widely employed as a prototypical model of stick--slip \cite{Rabinowicz_SciAm_1956}. The two solids in contact are approximated as rigid blocks, with the sliding force applied through a compliant spring. The friction force at the interface reduces with increasing velocity and increases with time (aging). Such a dependence is referred to as a rate-and-state law \cite{Dieterich_JGeophysRes_1979, Ruina_JGeophysRes_1983}. This model reproduces the main features of stick-slip behavior observed in metals, glassy polymers and rocks \cite{BaumbergerCaroli_AdvPhys_2006, Scholz_Nature_1998}, where the rigid body approximation is a reasonable one. 

However, the presence of significant elastic deformation, coupled with interfacial adhesion, invalidates this conceptual picture---interface deformation affects the sliding dynamics while adhesive forces cause intimate contact between the surfaces \cite{JohnsonETAL_ProcRoySocA_1971}. As a result, this type of contact, typical of adhesive polymer interfaces, does not conform well to rate-and-state laws \cite{BaumbergerCaroli_AdvPhys_2006}. Therefore, a new description of the stick--slip dynamics is required for these surfaces, accounting for both the interface deformation and the intimacy of contact.

An important example of adhesion and elasticity dominated sliding arises in low velocity sliding of soft polymers such as natural rubber. In these systems, stick--slip motion is known to result from the propagation of Schallamach waves \cite{Schallamach_Wear_1971, Barquins_Wear_1985, RandCrosby_ApplPhysLett_2006, FukahoriETAL_Wear_2010}. These waves are termed slow waves because they propagate at speeds much slower than those of elastic waves. Schallamach waves result from buckling of the polymer surface \cite{Barquins_Wear_1985}. The propagation of a single wave results in unit slip at the interface, which otherwise remains stationary between successive waves. Experimental studies of these waves, typically using a spherical indenter as one of the contacting solids, have helped establish several important features. These include the role of viscoelastic effects in buckling \cite{RandCrosby_ApplPhysLett_2006, BarquinsCourtel_Wear_1975, BestETAL_Wear_1981}, surface wrinkling \cite{KoudineBarquins_JAdhSciTech_1996}, critical nucleation velocity \cite{Schallamach_Wear_1971, Barquins_Wear_1985, MaegawaNakano_Wear_2010} and a quantitative analogy with elastic dislocations \cite{ViswanathanETAL_PhysRevE_2015}. 

Related recent work has suggested that slow waves might in themselves be a more general phenomenon, common to elastically deformable surfaces. Different types of slow waves have been reported in systems as varied as model tectonic faults \cite{NielsenETAL_GeophysIntJ_2010}, foam rubber blocks \cite{AnooshehpoorBrune_PureApplGeophys_1994}, poroelastic hydrogels \cite{BaumbergerETAL_PhysRevLett_2002} and polymeric glasses \cite{RubinsteinETAL_Nature_2004}. Theoretical models have also postulated the occurrence of slow waves on the basis of phenomenological friction force laws \cite{BarSinaiETAL_GeophysResLett_2012, DaubETAL_JGeophysRes_2010}. This raises a key question with regard to sliding of soft adhesive surfaces: are other types of stick--slip motions mediated by slow waves possible, over and above Schallamach waves? 

The present work seeks to investigate this question by building on observations of single stick--slip events in a unique adhesive contact geometry \cite{ViswanathanETAL_SoftMatter_2016}. It is shown, based on analysis of friction force traces that stick--slip itself is of three different types. High--speed \emph{in situ} imaging reveals that these correspond to the motion of either Schallamach waves or one of two new slow frictional waves---caled separation pulses and slip pulses. The \emph{in situ} imaging enables individual wave properties to be extracted quantitatively, in addition to establishing several important features of each of the waves.  

The manuscript is organized as follows. The experimental configuration used is described in Sec.~\ref{sec:experimental}. The results are presented, along with a discussion of their implications, in Sec.~\ref{sec:results}. In addition to describing wave properites, the importance of adhesion is also demonstrated. A summary of the observations is provided in Sec.~\ref{sec:summary}.

\section{Experimental}
\label{sec:experimental}

\begin{figure}
\centering
	\includegraphics[width=\textwidth]{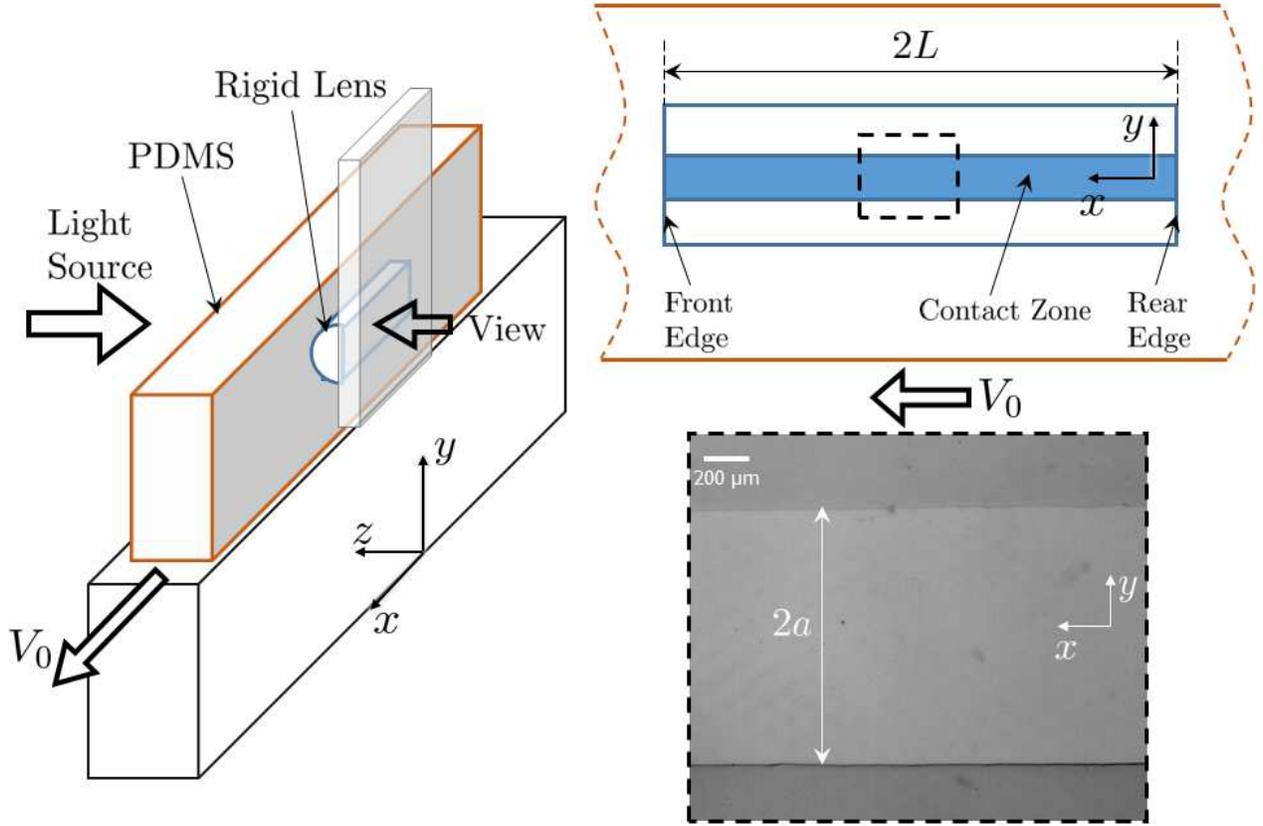}

	\caption{Experimental setup. (Left) An orthographic view of the experimental setup. The PDMS (brown) is attached to a rigid slide (black) and brought into contact with the rigid lens (blue). $\slideVel$ is applied to PDMS via the slide, remote from the interface. Coordinate system used in is also shown. (Right, top) Projection view of the contact region, as seen through the transparent lens. The front and rear contact edges are defined with respect to $\slideVel$, as shown here. (Right, bottom) A sample high-resolution image of the contact region (dashed lines) corresponding to the region marked in the figure at right, top (dashed lines). $2L, 2a$ are contact dimensions along $x,y$ directions respectively.}
	\label{fig:experimental}
\end{figure}

Quantitative information about interface dynamics is obtained by combined force measurements and high-speed \emph{in situ} imaging of a model adhesive interface. Figure~\ref{fig:experimental} shows a schematic of the experimental setup, involving intimate contact between a polydimethylsiloxane (PDMS, Dow Corning Sylgard 184) slab and an optically smooth plano-convex lens (Edmund Optics). 

The PDMS was formed by thoroughly combining a base (vinyl--terminated polydimethylsiloxane) with a curing agent (methylhydrosiloxane--dimethylsiloxane copolymer) in the weight ratio 10:1. The mixture was allowed to cure in a rectangular mold with dimensions of 22 mm $\times$ 70 mm (sliding length) in the $xy$ plane (see Fig.~\ref{fig:experimental}). Curing time was fixed at 6 hours at 100$^\circ$C, followed by 18 hours at room temperature ($\sim 25^\circ$ C). One side of the mold was covered by a polycarbonate sheet, and this surface was used for sliding experiments. For experiments involving Schallamach waves, a longer sliding of 90 mm was used. The thickness of the slabs was varied between 20 mm and 30 mm with no changes in the observed dynamics. The Young's modulus and Poisson's ratio for PDMS were estimated as 1 MPa and 0.46 respectively, from shear \cite{Mark_PolymerDataHandbook_2009} and bulk modulus values reported in the literature \cite{SchmidMichel_Macromolecules_2000}. 

The lens has a cylindrical plano-convex geometry with face radius $16.25$ mm. Upon contacting the flat PDMS surface, the lens formed a very long adhesive contact of length $2L = 25$ mm and constant contact width $2a\, (\ll 2L)$. In contrast to the commonly used spherical contact geometry, this cylindrical configuration allowed the isolation of individual stick--slip events, without any edge effects.

The contact interface was illuminated by a uniform backlight source (Metaphase Technologies). High--speed, high--resolution \emph{in situ} imaging of the interface was enabled by a microscope (Nikon Optiphot) coupled to a high--speed CMOS camera (PCO dimax). Depending on the particular experimental run, frame rates between 100 and 5000 fps were used, with a resolution of $2.8$ $\mu$m per pixel. The high-speed image sequences obtained from the experiments were analyzed differently depending on the type of wave being studied. These analysis procedures are described in detail in Ref.~\cite{ViswanathanETAL_SoftMatter_2016}.

The PDMS was attached to a linear slide capable of providing constant remote velocities in the range of $10$ $\mu$m/s to $20$ mm/s. The camera and the lens were fixed on rigid supports as shown in Fig.~\ref{fig:experimental} (left). Simultaneous with the \emph{in situ} observations, the friction force was measured using a piezoelectric dynamometer (Kistler Model 9254). \markthis{The dynamometer was attached to the cylindrical lens, mounted on a rigid support. Mounting the dynamometer onto the lens did not change the system stiffness due to the very high compliance of the polymer-lens interface.} The force measurements were correlated with the \emph{in situ} imaging to outline the interface dynamics.

In a typical experiment, the PDMS and lens were first brought into contact with a fixed $2a$. This was followed by a wait time of $t = 60$ s, in order to eliminate effects of contact aging. \markthis{The loading in the $z$-direction was hence one of constant displacement. This enabled different wave propagation regimes to be delineated easily.} The remote velocity $\slideVel$ was then applied to the PDMS so that the resulting sliding distance ($x$-direction) was at least 30 mm. The effect of different normal forces was quantified in terms of a change in $2a$. Hence the two experimental parameters were $2a$ and $\slideVel$.

\section{Results and Discussion}
\label{sec:results}

The sliding behavior of the model adhesive contact was studied over a range of velocities and normal loads (or $2a$). Stick-slip was observed to be the primary mode of sliding for this range of $\slideVel$ (0.05 to 10 mm/s). However, distinct \lq modes\rq\ of stick-slip were observed and the associated interface motion was found to depend on the values of $2a$ and $\slideVel$. 

\subsection{Three modes of stick--slip}

Conventionally, stick-slip motion has been identified with two independent properties of the friction force \cite{BowdenTabor_Friction_1973} --- oscillation at a constant frequency and approximately constant force reduction in each oscillation cycle. Any friction force trace that obeys these properties is usually attributed to stick--slip motion.

\begin{figure}
  \centering
  \mbox
  {
    \subfigure[\label{fig:SchW_forceTrace}]{\includegraphics[width=0.5\textwidth]{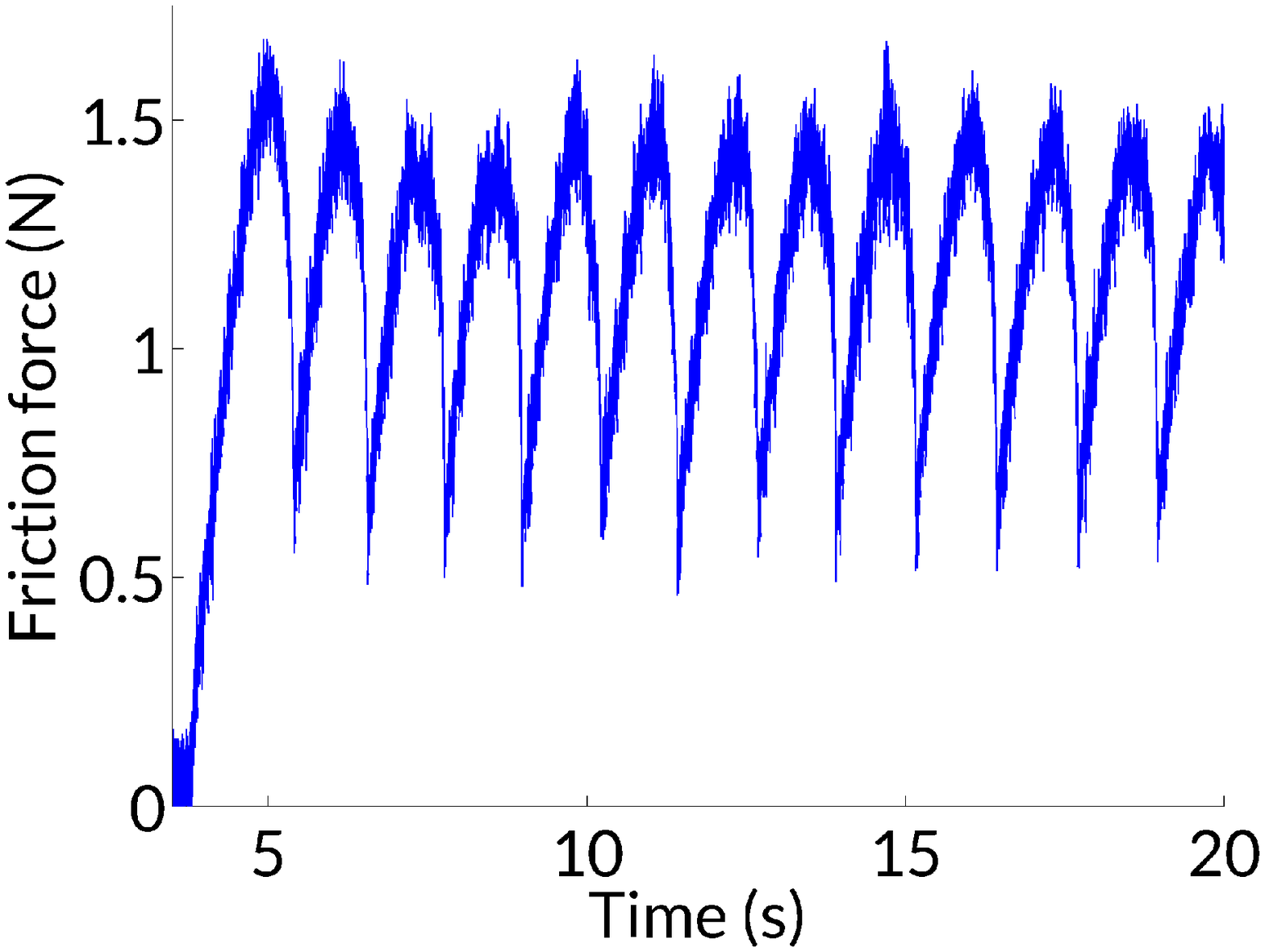}}\quad
    \subfigure[\label{fig:SP_forceTrace}]{\includegraphics[width=0.5\textwidth]{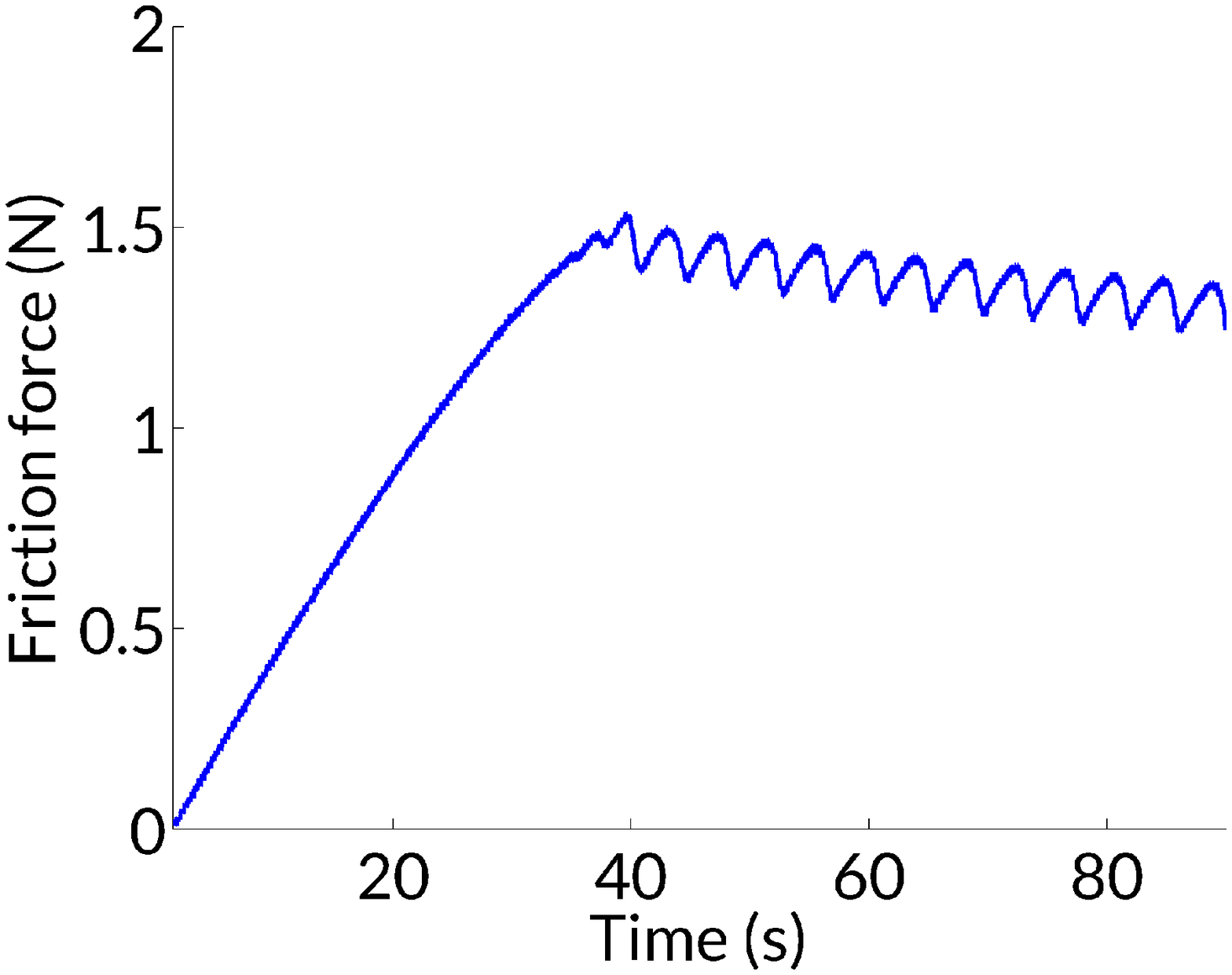}}


  }\\
  \mbox
  {
    \subfigure[\label{fig:DP_forceTrace}]{\includegraphics[width=0.5\textwidth]{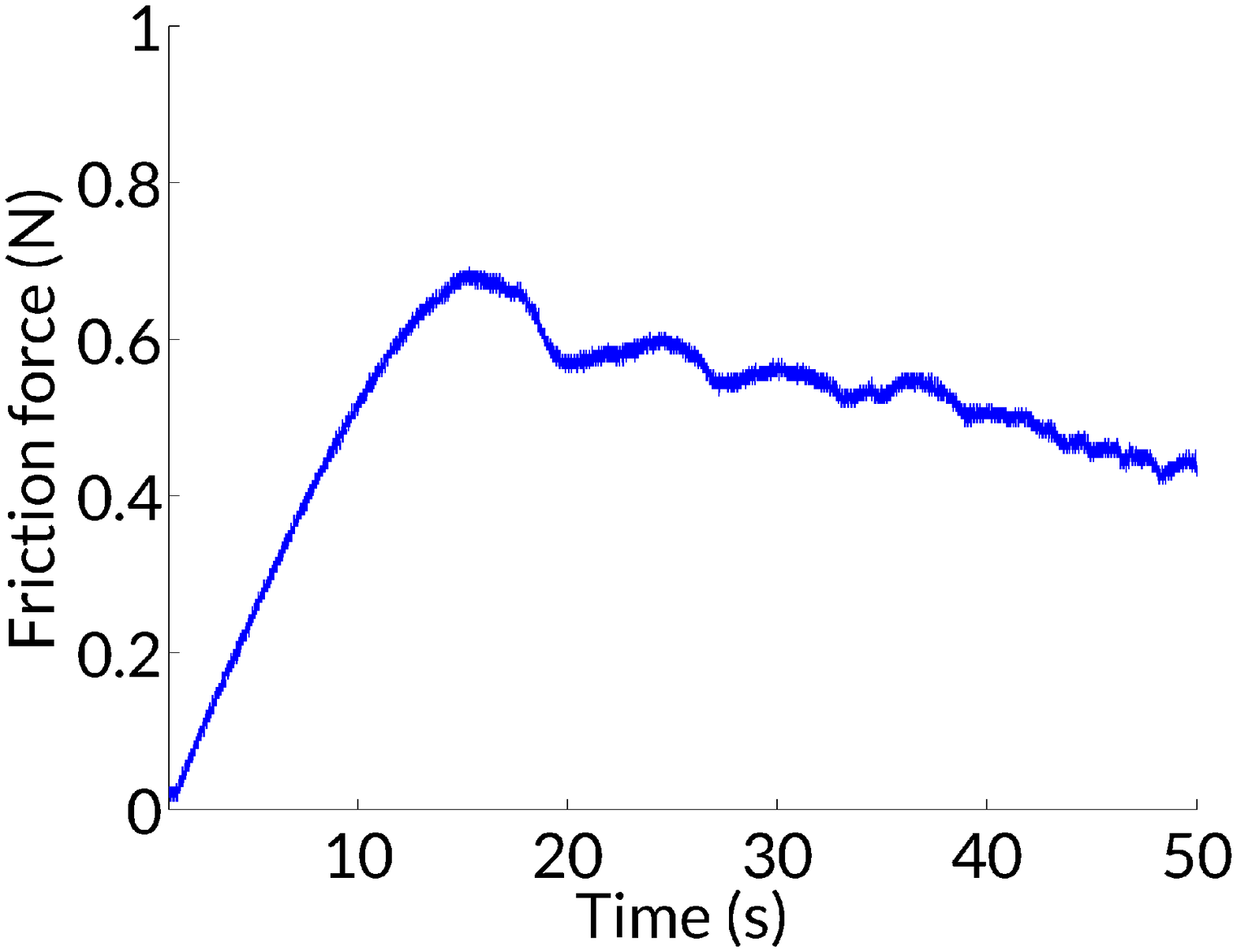}}
  }
  \caption{Three distinct force traces showing \lq stick-slip\rq\ characteristics. (a) For $\slideVel > V_C$ a critical sliding velocity, high oscillation frequency ($\sim 1$ Hz) is observed with large force reduction in each cycle ($\sim 1$ N). (b) At $\slideVel < V_C$, but $2a= 2100 \mu$m), the frequency of oscillation and force amplitude reduce to $\sim 0.19$ Hz and 0.2 N respectively. (c) Same condition as in (b) but for smaller load $2a = 900 \mu$m. Now the frequency of oscillation reduces even further (0.17 Hz) and force reduction per cycle is only $\sim 0.1$ N. The force traces provide evidence for three different types of stick-slip. Data in (a) is adapted from Ref.~\cite{ViswanathanETAL_SoftMatter_2016}.}
  \label{fig:forceTraces}
\end{figure}

In the case of adhesive elastic contacts, a surprisingly different picture emerged from the friction force measurements. Three distinct types of force traces were observed, depending on the values of $\slideVel$ and $2a$ in the sliding, see Fig.~\ref{fig:forceTraces}. It is clear that the three types have characteristic oscillation frequencies as well as force reduction values. They are all different, yet can all be classified as arising due to \lq stick-slip.\rq 

For $\slideVel $ greater than a critical value $V_C \sim 200\,\mu$m/s, large amplitude force reductions ($\sim 1$ N) were observed, see Fig.~\ref{fig:SchW_forceTrace}. The force trace in Fig.~\ref{fig:SchW_forceTrace} clearly shows two phases --- a stick phase where the force increases to a maximum value, and a slip phase when a rapid force reduction occurs. The oscillation frequency in this $\slideVel$ range ($n\simeq 1$ Hz) was found to be linearly proportional to $\slideVel$. The peak force reduced by a few percent between consecutive cycles but the frequency, for a given $\slideVel$, was constant.

When $\slideVel  < V_C$, two different situations arose, depending on the value of $2a$. For $2a > 1500\,\mu$m, the force reduction per cycle and oscillation frequency were significantly reduced (Fig.~\ref{fig:SP_forceTrace}), with corresponding values of 0.2 N and $n = 0.19$ Hz, respectively. Again, the oscillation frequency was constant, although some variation in peak force was observed between cycles.

For $\slideVel < V_C$ and $2a < 1500\, \mu$m, a third type of force trace was observed, shown in Fig.~\ref{fig:DP_forceTrace}. The friction force showed an additional reduction in amplitude $(<0.1$ N) and frequency ($n\sim 0.17$ Hz). As with the other two traces, $n=$const. but the force reduction per cycle varied from one cycle to the next. The mean friction force was lower as well (\emph{cf.} Fig.~\ref{fig:forceTraces} (top row)), due to the reduced contact width $2a$. 

The three distinct force traces of Fig.~\ref{fig:forceTraces} were observed under different $\slideVel, 2a$ conditions. They correspond to fundamentally different types of stick--slip motion, and are henceforth referred to as stick--slip modes. The origin of these three modes cannot be ascertained based on force measurements alone, but was instead revealed by observing the interface \emph{in situ} during sliding.

\subsection{\emph{In situ} observations of interface dynamics}

While force traces provide average information about the sliding process, they are fundamentally incapable of revealing the mechanisms behind the three observed stick-slip modes. This information was obtained from direct high--speed \emph{in situ} imaging (Sec.~\ref{sec:experimental}) of the contact during sliding.

The \emph{in situ} observations revealed that the three stick-slip force traces, Figs.~\ref{fig:SchW_forceTrace},~\ref{fig:SP_forceTrace} and Fig.~\ref{fig:DP_forceTrace} were caused by the passage of three distinct types of waves at the interface, called the Schallamach wave, slip pulse and separation pulse, respectively. In each case, the interface moved by a unit distance due to the propagation of a single wave and remained stationary between successive waves. Of the three different waves observed, two---the separation pulse and the well-known Schallamach wave---involved local interface detachment. The third wave, the slip pulse, was manifest as a localized stress front. 

\begin{figure}
\centering
	\includegraphics[width=\textwidth]{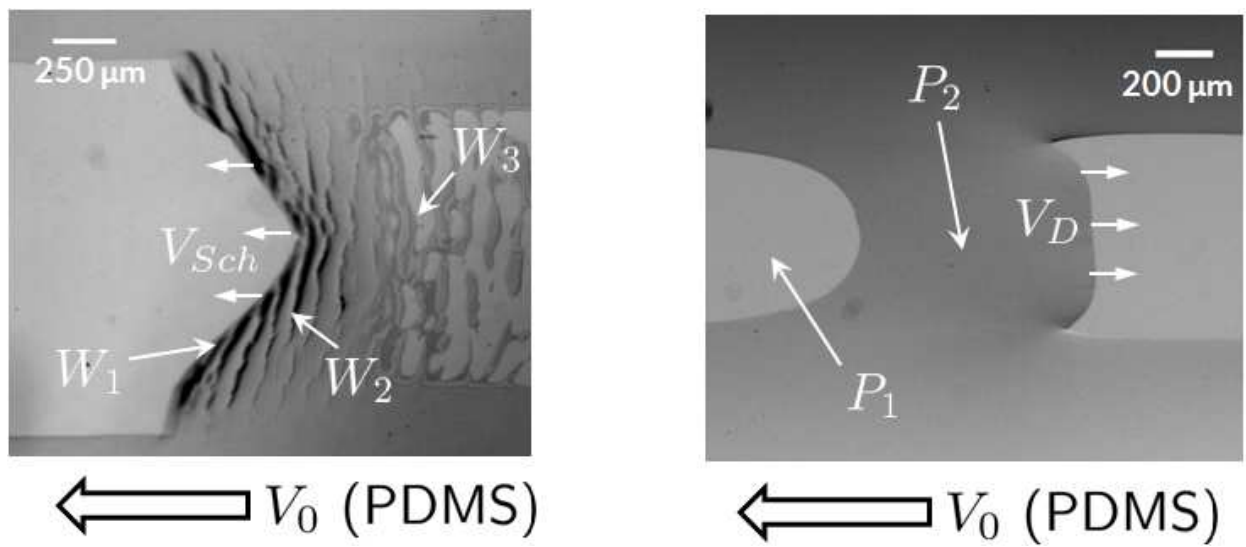}
	\caption{Opposite propagating interface waves with detachment. (Left) Frame showing a single Schallamach wave propagating at speed $V_{Sch}$ in the contact. The wave consists of a distinct detached region (front at $W_1$) where the PDMS is not in contact with the lens. The wrinkles ($W_2$) accompanying the wave cause incomplete adhesion in after wave passage ($W_3$). (Right) Frame showing separation pulse propagation. Like the Schallamach wave, this wave is also a local region of detachment ($P_2$). However, in contrast to the Schallamach wave, the separation pulse travels in a direction opposite to $\slideVel$, with velocity $V_D \ll V_{Sch}$. Perfect adhesive contact is re-established after wave passage ($P_1$).}
	\label{fig:dwave_frames}
\end{figure}

Frames from a high-speed sequence depicting the propagation of a single Schallamach wave and a separation pulse are shown side-by-side in Fig.~\ref{fig:dwave_frames}. These waves were responsible for the force traces of Fig.~\ref{fig:SchW_forceTrace} and~\ref{fig:DP_forceTrace}, respectively. The Schallamach wave (Fig.~\ref{fig:dwave_frames}, left) is seen as a local region of detachment (front at $W_1$) that propagates in the same direction as $\slideVel$. Within this wave, the PDMS and lens temporarily lose contact with one another and become readhered after wave passage. However, the re-formed contact is littered with air pockets (arrow at $W_3$). These are due to the wrinkles accompanying a Schallamach wave that act as strain concentrators and prevent complete interface relaxation \cite{ViswanathanETAL_PhysRevE_2015}. The velocity $V_{Sch}$ of Schallamach wave propagation was constant and a function of $\slideVel$---for the range considered, $V_{Sch}/\slideVel  = 50 \text{ to } 100$. Between successive Schallamach waves, the interface remained stationary, corresponding to the stick-phase of the stick-slip cycle. The number of Schallamach waves per second was equal to the frequency of force oscillation ($n=1$ Hz, see Fig.~\ref{fig:SchW_forceTrace}).

The separation pulse, though also a local region of detachment like the Schallamach wave, has fundamentally distinct properties, see Fig.~\ref{fig:dwave_frames} (right). Firstly, it propagates opposite in direction to the Schallamach wave (and so opposite in direction to $\slideVel$ also) at a much lower velocity $V_D \leq 20 \slideVel$. Secondly, it is observed only at low applied normal loads $(2a)$. Thirdly, in contrast to the Schallamach wave, the detachment zone $P_2$ of a separation pulse is devoid of any compression-induced features such as wrinkles. As a result, perfect readhesion is observed after single wave passage ($P_1$). It is noteworthy that even though separation pulses and Schallamach waves propagate in opposite directions, the interface always slips in the same direction (parallel to $\slideVel$). Yet again, as with the Schallamach waves, the number of separation pulses per second was found to equal the force oscillation frequency. Between successive pulses, the interface remained stationary (the stick-phase).

\begin{figure}
\centering
	\includegraphics[width=0.6\textwidth]{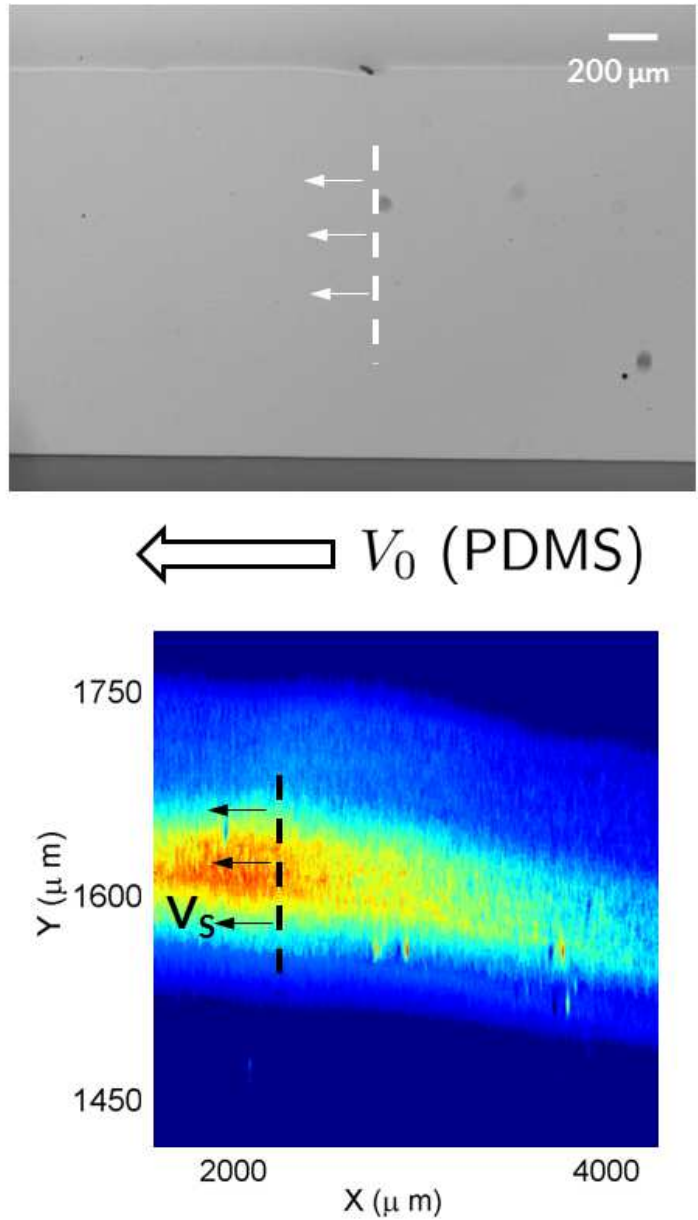}
	\caption{Slip pulse propagation. (Top row) In the optical image, dirt particles inside the contact are seen to move in the direction of $\slideVel$. However, wave features are not visible since interface detachment does not occur. (Bottom row) False-color image of shear stress in part of the contact, obtained using cross polarizers. Interface motion is seen to be effected by a slip pulse in the form of a local stress front (dashed arrows) moving at velocity $V_S \gg \slideVel$. Red and blue correspond to maximum and minimum shear stress (arbitrary units).}
	\label{fig:SP_frames}
\end{figure}

The third wave---the slip pulse---was found to cause the force trace in Fig.~\ref{fig:SP_forceTrace}. A slip pulse does not involve any interface separation at all. For this reason, slip pulses are difficult to image since no visual interface features exist for tracking purposes. This scenario is illustrated in Fig.~\ref{fig:SP_frames}. In the optical image (top row), the interface is seen to move locally during the slip phase, and then become arrested at the onset of the stick phase. This cycle caused interface displacement to occur in steps, as confirmed by observing the motion of tracer particles on the PDMS surface. However, when imaging using crossed polarizers, the interface shear stresses are revealed directly \cite{ViswanathanETAL_SoftMatter_2016}, see Fig.~\ref{fig:SP_frames} (bottom row). Now the slip cycle is seen to be initiated by the propagation of a single stress front (marked by white dashed line) at a constant speed $V_S \simeq 400 \slideVel$. As seen in the color image, this front is preceded by a large stress build-up during the stick phase that quickly relaxes in the wake of the pulse. From this image, specific details about unit interface slip as well as local pulse velocity were obtained by suitable processing routines \cite{ViswanathanETAL_SoftMatter_2016}. Just as with Schallamach waves and separation pulses, the number of slip pulses per second was the same as the force oscillation frequency (Fig.~\ref{fig:SP_forceTrace}).

\subsection{Wave parameters and nucleation}

The primary wave parameters were found to be the wave speed $\tilde{V} = V_D, V_S, V_{Sch}$, the unit slip per wave $\Delta x =\Delta x_D, \Delta x_S, \Delta x_{Sch}$ and the frequency of oscillation $n =n_D, n_S, n_{Sch}$. The subscripts $D, S, Sch$ are used to denote separation pulses, slip pulses and Schallamach waves, respectively. These parameters are fixed during nucleation and uniquely determine the properties of the waves. 

Wave observations at different $\slideVel$ and $2a$ revealed that $\Delta x$ and $n$ were related to the sliding velocity $\slideVel$ by
\begin{equation}
\label{eqn:ndx}
\slideVel = n \Delta x
\end{equation}

This relation was found to hold within a few percent for Schallamach waves, separation pulses and slip pulses. Therefore, it explicitly establishes that interface motion arises solely due to wave propagation at the interface. The implications of this equation, however, are slightly different for each of the three waves. 

\begin{figure}
  \centering
  \mbox
  {
    \subfigure[\label{fig:DP_genFreq}]{\includegraphics[width=0.6\textwidth]{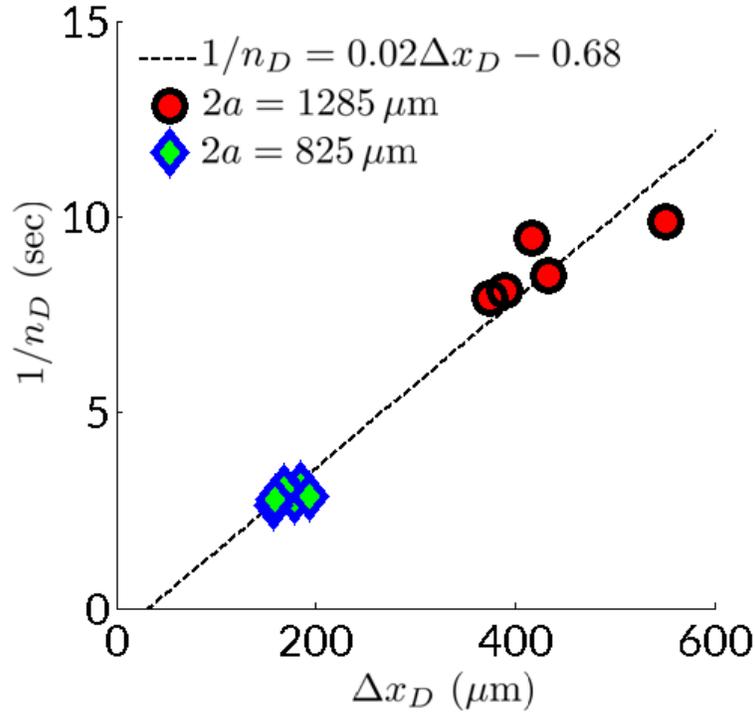}}
  }\\
  \mbox
  {
    \subfigure[\label{fig:SchW_genFreq}]{\includegraphics[width=0.65\textwidth]{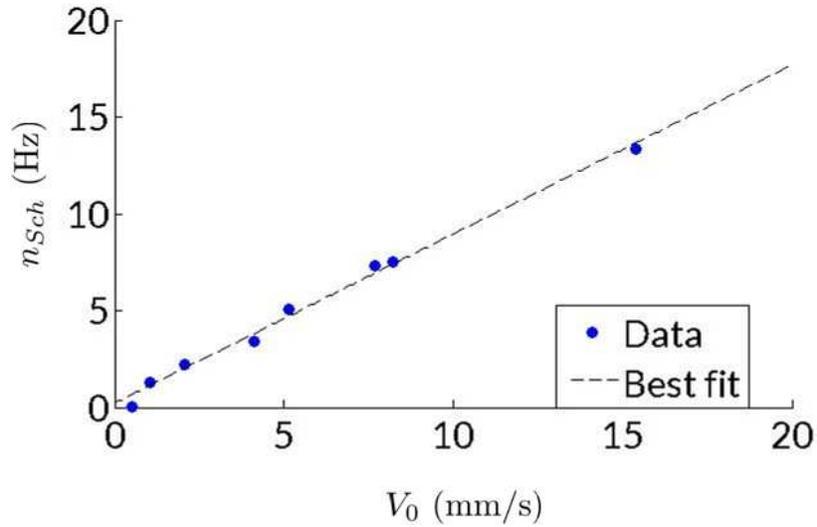}}
  }
  \caption{Relation between frequency and unit slip for (a) separation pulses and (b) Schallamach waves. The relation $\slideVel = n \Delta x$ holds for both waves. However, for separation pulses, $\Delta x_D$ changes with $2a$. \markthis{The best fit line (dashed) has inverse slope equal to $\slideVel = 50\,\mu$m/s}. For Schallamach waves, $\Delta x_{Sch} = $ const. \markthis{(inverse of slope of best-fit line)} over a large $\slideVel$ range. }
  \label{fig:genFreqs}
\end{figure}

Firstly, for separation pulses, the amount of slip $\Delta x_D$ was found to depend on both $\slideVel$ and $2a$, see Fig.~\ref{fig:DP_genFreq}. Correspondingly, it is seen in the figure that $n_D$ also depends on $2a$ and $\slideVel$ so that their product obeys Eq.~\ref{eqn:ndx} to within $6\%$. In fact, the best fit line (dashed in the figure) has inverse slope nearly equal to $\slideVel = 50\,\mu$m/s.  

Secondly, for Schallamach waves, it is known that the slip $\Delta x_{Sch}$ is constant, independent of $\slideVel$ \cite{ViswanathanETAL_PhysRevE_2015}. This value is determined by the extent of the buckling zone on the PDMS surface. Now Eq.~\ref{eqn:ndx} is still obeyed but $n_{Sch}$ is proportional to $\slideVel$ instead, see Fig.~\ref{fig:SchW_genFreq}. In fact this proportionality holds over a large range of $\slideVel$ values, as seen from the best fit line (dashed) with slope equal to $\Delta x_{Sch}$. \markthis{An interesting feature is that the line has non-zero intercept on the $n_{Sch}$ axis---finer observations revealed that the $n$-$\slideVel$ relationship was non-linear as $\slideVel \to V_C$}.

Finally, for slip pulses, the $\Delta x_S$, $n_S$ dependence on $\slideVel$ was found to be similar to that of separation pulses. In this case, Eq.~\ref{eqn:ndx} was found to hold to a much larger accuracy (less than 2\% over several experimental runs). 

\subsection{Simultaneous waves and the separation--slip pulse transition}

Schallamach waves are nucleated by a buckling instability on the PDMS surface, while separation pulses are nucleated by a tensile peel off process. These two nucleation events are independent and can occur simultaneously. Consequently, in several experiments with $\slideVel > V_C$, both separation pulses and Schallamach waves were observed in the same contact region. As the two waves propagated in opposite directions, they collided with each other, resulting in a single stagnant air pocket. During this interaction, the interface remained stationary at the point of contact; subsequently another Schallamach wave was nucleated at the stagnant air pocket and continued its propagation inside the contact region. 

Separation pulses and slip pulses occur at the same $\slideVel$ but at different $2a$ values. A transition between separation and slip pulses was observed by continuously changing $2a$ during an experimental run. The force was found to transition from that characteristic of separation pulses to that of slip pulses beyond a critical $2a$ value. Correspondingly, separation pulse propagation was arrested beyond this critical value and interface motion was mediated by slip pulses instead. Details of this transition, including a high-speed movie, are presented in Ref. \cite{ViswanathanETAL_SoftMatter_2016}. 

\begin{figure}
\centering
	\includegraphics[width=0.6\textwidth]{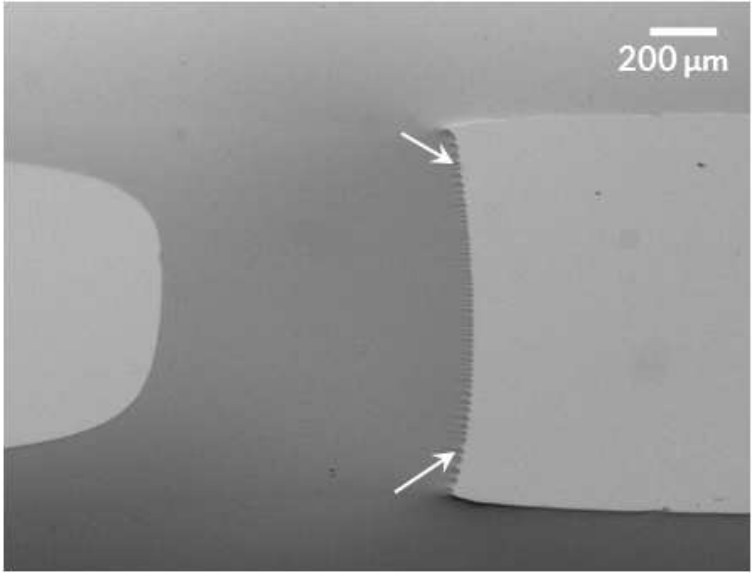}
	\caption{Shape of separation pulse near the $2a$ transition. Large tensile strains are evident in the wrinkle marks (at arrows) at the front of the wave. For slightly larger $2a$ values, detachment is not observed, and a slip pulse propagates instead.}
  \label{fig:transition}
\end{figure}

It is interesting to note the shape of a separation pulse as one approaches the critical $2a$ value, see Fig.~\ref{fig:transition}. This represents the physical limit at which the PDMS surface can locally detach from the lens. Consequently the head of the detached region shows tensile wrinkles, resembling stretch marks in plastic sheets. These marks were nearly absent for separation pulses at lower $2a$ values (\emph{cf.} Fig.~\ref{fig:dwave_frames} (right)). It is likely that the size of these wrinkles continuously increases with $2a$ close to the transition point. At the transition, it is energetically less favorable for the PDMS to detach locally. Instead, the PDMS slips locally in the form of a slip pulse, without any detachment.

\subsection{Importance of intimate adhesive contact}

\begin{figure}
\centering
	\includegraphics[width=0.5\textwidth]{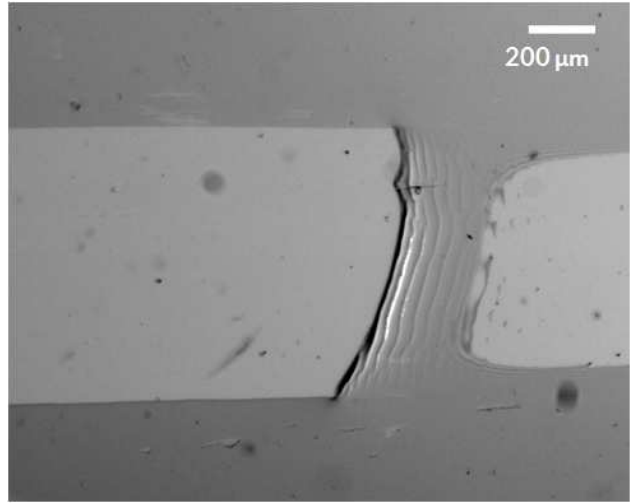}
	\caption{Schallamach wave propagation in an imperfect contact. Despite the presence of dirt particles/ scratches inside the contact region, Schallamach wave propagation occurs albeit with altered geometry as shown here. The wave shape is distinctly different from a single wave inside a perfect adhesive contact with no impurities. }
	\label{fig:dirtyContact}
\end{figure}

The generality of the wave observations was studied using imperfect adhesive contacts. It was observed that Schallamach waves still propagated in contacts with significant impurities. Figure~\ref{fig:dirtyContact} shows an image of a Schallamach wave in a contact with a large number of dirt particles and scratches. When compared to Schallamach waves in a perfect adhesive contact (Fig.~\ref{fig:dwave_frames} (left)), certain distinct differences may be observed. Firstly, the shape of the wave is no longer symmetric with respect to the contact mid-line ($y=0$ in Fig.~\ref{fig:experimental}). The wrinkles accompanying the wave are also seen to be more even and spread out. Secondly, following wave passage, air pockets are left behind over single dirt particles inside the contact. These are reminiscent of dislocation loops left behind when dislocations interact with second phase particles in a glide plane \cite{ViswanathanETAL_PhysRevE_2015}. 

In contrast to Schallamach waves, slip pulses and separation pulses were found to be sensitive to the presence of contact impurities. In the absence of a clean adhesive contact, separation pulse propagation was no longer observed, even though nucleation events occurred at multiple local sites in the interface. Hence, the presence of a JKR-like adhesive contact \cite{JohnsonETAL_ProcRoySocA_1971} was found to be essential for effecting wave propagation.

\begin{figure}
\centering
	\includegraphics[width=\textwidth]{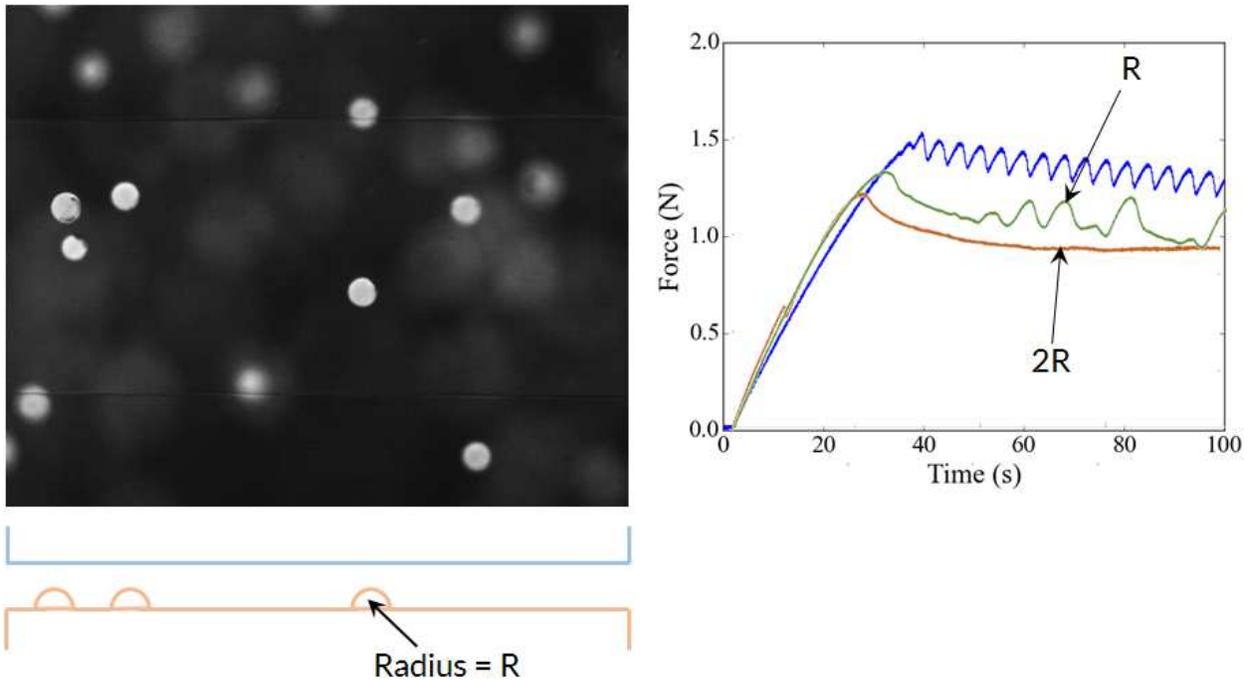}
	\caption{Role of adhesion in frictional wave propagation. (Left) Polarized dark field image showing fluorescent polyethylene spheres inside the contact. Spheres in focus are located on the PDMS surface, as depicted in the schematic. (Right) Force trace with spheres of radius $R = 50\,\mu$m and $2R = 100\,\mu$m shows that slip pulse propagation becomes increasingly difficult as the contact transitions from being adhesive (JKR) to Hertzian elastic.}
  \label{fig:polyethylene}
\end{figure}

To confirm the role of adhesion, a controlled experiment was performed to simulate the transition from intimate adhesive contact to a Hertz-like elastic one. Prior to curing, the PDMS was thoroughly mixed with fluorescent monodisperse polyethylene spheres of different radii (Radius = $R = 50\,\mu$m and Radius = $2R = 100\,\mu$m). The amount of polyethylene introduced was fixed to ensure that, after homogeneous mixing, several spheres were protruding on the sliding surface, see Fig.~\ref{fig:polyethylene} (left). In the polarized light image, the PDMS appears dark due to the crossed polarizers used. On the other hand, the polyethylene spheres, being fluorescent, are seen as bright circles. Spheres on the PDMS surface in contact with the lens are seen to be in focus while those in the bulk are more diffuse. Hence at the contact, the normal load was mostly transmitted to the protruding spheres, preventing the formation of a pristine intimate adhesive contact (see schematic in Fig.~\ref{fig:polyethylene} (left))

Three PDMS samples were used --- one with spheres of radii $R$, one with spheres of radii $2R$ and the other without any spheres (plain PDMS). Testing conditions were chosen corresponding to passage of slip pulses in the plain PDMS ($2a = 2100\,\mu$m, $\slideVel = 50\,\mu$m/s). The corresponding force traces are shown in Fig.~\ref{fig:polyethylene} (right). When the plain PDMS was slid against the lens, the force trace showed stick-slip characteristics (Fig.~\ref{fig:polyethylene}, right, blue curve), consistent with slip pulse propagation (see Fig.~\ref{fig:SP_forceTrace}). The PDMS with spheres of radii $R$ showed some signs of stick-slip but with homogeneous sliding in between (green curve). In stark contrast, stick-slip features were completely absent in the PDMS with $2R$ spheres (orange curve) reflecting purely homogeneous motion. 

These observations are consistent with the hypothesis that an intimate adhesive contact is necessary to effect slip/ separation pulse propagation. When spheres with radius $R$ are in contact with the lens, small domains of adhesive contact remain, allowing partial slip pulse propagation. This is consistent with the odd stick-slip cycle observed in the force trace. However, for the larger spheres, these adhesive contact domains are largely absent and homogeneous sliding ensues, as one would expect for purely elastic materials without adhesion. 

\subsection{Mechanics of wave propagation}

\begin{figure}
\centering
	\includegraphics[width=\textwidth]{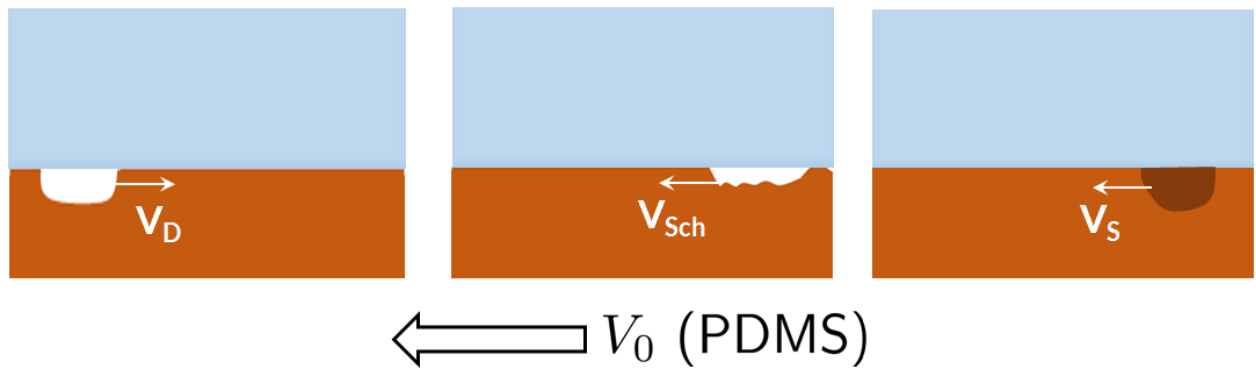}
	\caption{Schematic side view of contact showing the three slow frictional waves. White arrows indicate wave propagation direction. (Left) The separation pulse is seen as a local region of detachment moving in a direction opposite to $\slideVel$. (Middle) The Schallamach wave is also a local detachment region, but it moves in the same direction as $\slideVel$. (Right) A slip pulse is a compressive stress (dark brown) front that propagates in the same direction as $\slideVel$ with no detachment at the interface. }
	\label{fig:schematic}
\end{figure}

It is evident from the \emph{in situ} observations that the interface dynamics is quite different for each of the observed stick-slip modes. A schematic side-view of the interface is presented in Fig.~\ref{fig:schematic}, depicting the propagation of the corresponding slow frictional waves. The separation pulse and Schallamach wave, both involving local interface separation, were comprised of a trapped air pocket that traversed the contact (Fig.~\ref{fig:schematic}, left and middle). As noted earlier, Schallamach waves propagated in the same direction as $\slideVel$, while separation pulses propagated in the opposite direction. The slip pulse, on the other hand, is a local compressive surface zone that moved in the same direction as $\slideVel$ (Fig.~\ref{fig:schematic}, right). 

The direction of wave propagation vis-\'a-vis $\slideVel$ can be easily explained in terms of the tension/ compression at the free surface \cite{ViswanathanETAL_SoftMatter_2016}. Using the coordinate system shown in Fig.~\ref{fig:experimental}, the direction of propagation is linked to the sign of the surface strain $\epsilon_{xx}$. For steady-state constant velocity ($\tilde{V} = V_{Sch}, V_S$ or $V_D$) propagation, the tangential interface displacement $u_x$ is a function $u_x (x - \tilde{V} t)$. The surface strain $\epsilon_{xx}$ can be approximated as
\begin{equation}
	\epsilon_{xx} = \frac{\partial u_x}{\partial x} \simeq -\tilde{V}^{-1} \frac{\Delta x}{T}
\end{equation}
where $\Delta x$ is the slip per wave and $T$ is duration of propagation. The sign of $\Delta x$ is the same as $\slideVel$, so that the strain $\epsilon_{xx}$ is negative (compressive) if the wave propagates in the same direction as $\slideVel$ and is positive (tensile) if it propagates in the opposite direction. 

Therefore, for the case Schallamach waves and slip pulses $\epsilon_{xx} <0$, and they result in compressive surface strain. For separation pulses $\epsilon_{xx} > 0$ and the surface strain is tensile. 

The propagation phase of all three waves can be described theoretically using a common elastodynamic framework \cite{ViswanathanETAL_SoftMatter_2016_2}. This theoretical treatment not only reproduces the qualititative features of wave propagation (direction, frequency dependence on $\slideVel$) but also provides closed-form solutions for interface stresses, displacements and velocities accompanying wave propagation.  

\section{Summary}
\label{sec:summary}

The coupled \emph{in situ} imaging and force measurements revealed that the notion of stick-slip in adhesive polymer interfaces is more general than previously thought. A detailed analysis of the accompanying interface dynamics showed that three modes of stick-slip may be uniquely identified in low velocity sliding. Each of these modes was found to comprise of cyclical stick and slip phases, with distinct force traces. Correspondingly, the interface slip in each cycle was caused by the propagation of a single wave, with the type of wave determining the stick-slip mode observed. The properties of the three waves---Schallamach waves, separation pulses and slip pulses---were described along with conditions favoring their propagation. The important role of adhesion in effecting the wave phenomena was also established.

\section*{Acknowledgments}
The authors would like to gratefully acknowledge the use of experimental facilities at the Center for Materials Processing and Tribology at Purdue Univesity. This research was supported in part by the U.S. Army Research Office via Award W911NF-15-1-0591, and NSF Grants CMMI 1562470 and DMR 1610094. K. V. would like to acknowledge financial support via the Bilsland Dissertation Fellowship at Purdue University.


\end{document}